\documentclass[10pt,journal,compsoc]{IEEEtran}
\ifCLASSOPTIONcompsoc
  \usepackage[nocompress]{cite}
\else
  \usepackage{cite}
\fi

\ifCLASSINFOpdf
\else
\fi

\usepackage{cite}
\usepackage{graphicx}
\usepackage[justification=centering]{caption}
\newcommand\scalemath[2]{\scalebox{#1}{\mbox{\ensuremath{\displaystyle #2}}}}
\usepackage{multirow}
\usepackage{array}
\usepackage{amsfonts}
\usepackage[ruled,linesnumbered]{algorithm2e}
\usepackage{float}
\usepackage{amsmath}
\newtheorem{myDef}{Definition}
\usepackage{booktabs}
\usepackage{amssymb}
\usepackage[utf8]{inputenc}
\usepackage{hyperref}
\usepackage{multicol}
\usepackage{adjustbox}
\usepackage{siunitx}
\usepackage{mhchem}
\usepackage{subfigure}

\newcommand{\etal}{\emph{et~al. }} 
\usepackage{textcomp}

\begin{document}

\bibliographystyle{IEEEtran}

\title{Higher-order Graph Attention Network for Stock Selection with Joint Analysis}

\author{Yang Qiao,
        Yiping Xia,
        Xiang Li, \textit{Member, IEEE,}
        Zheng Li,
        Yan Ge \textsuperscript{*}
\IEEEcompsocitemizethanks{
\IEEEcompsocthanksitem Qiao Yang and Yiping Xia are with Huawei Device Co., Ltd, Dongguan, Guangdong, 523808, China. 
 E-mail: \{qiaoyang19, xiayiping\}@huawei.com
\IEEEcompsocthanksitem Xiang Li and Yan Ge are with the Department of Computer Science, University of Bristol, Bristol, BS8 1QU, UK.\protect
E-mail: \{xiang.li, yan.ge\}@bristol.ac.uk
\IEEEcompsocthanksitem Zheng Li is with Zhengzhou University, 100 Kexue Blvd, Zhengzhou, Henan, 450001, China. 
E-mail: jiabaxia1@gmail.com\\
* Corresponding author. 
}\protect\\
}


\IEEEtitleabstractindextext{%
\begin{abstract}
Stock selection is important for investors to construct profitable portfolios. Graph neural networks (GNNs) are increasingly attracting researchers for stock prediction due to their strong ability of relation modelling and generalisation. However, the existing GNN methods only focus on simple pairwise stock relation and do not capture complex higher-order structures modelling relations more than two nodes. In addition, they only consider factors of technical analysis and overlook factors of fundamental analysis that can affect the stock trend significantly. Motivated by them, we propose \underline{\textbf{h}}igher-order \underline{\textbf{g}}raph \underline{\textbf{at}}tention network with joint analysis (H-GAT). H-GAT is able to capture higher-order structures and jointly incorporate factors of fundamental analysis with factors of technical analysis. Specifically, the sequential layer of H-GAT take both types of factors as the input of a long-short term memory model. The relation embedding layer of H-GAT constructs a higher-order graph and learn node embedding with GAT. We then predict the ranks of stock return. Extensive experiments demonstrate the superiority of our H-GAT method on the profitability test and Sharp ratio over both NSDAQ and NYSE datasets.




\end{abstract}

}

\maketitle

\IEEEdisplaynontitleabstractindextext

%
\IEEEpeerreviewmaketitle

\IEEEraisesectionheading{\section{Introduction}\label{sec:introduction}}

 \IEEEPARstart{S}{tock} market trading is a symbol of market capitalism. In 2022 the total market cap of the world’s stock market has exceeded \$105 trillion~\cite{statistaresearch}. In the field of quantitative finance, investing in high-growth potential stocks is a crucial topic. 
 However, making the right investment decisions and designing profitable trading strategies has many challenges due to the market’s highly volatile and non-stationary nature \cite{adam2016stock}. 
 
 Recent advances in deep learning present a promising prospect in stock forecasting.
For example, recurrent neural networks (RNNs), have become a promising solution to substitute the traditional time-series models to predict the future trend or exact price of a stock. However, most existing neural works treat stock movements as independent of each other, which is contrary to true market function, and makes forecasting challenging \cite{zhang2017stock, zhao2017constructing}. In reality, stocks are often related to each other, and there exist rich signals in relationships between stocks (or companies) \cite{nobi2014effects}. Recent advances in graph-based deep learning \cite{wu2020comprehensive} have led to the rise of graph neural networks (GNNs) that can model the relationships between related stocks. A stock prediction task can be formulated as a node classification task, where each node represents a stock and edges represent relations (e.g., the same CEO or belonging to the same industry) between companies. Graph Convolution Network (GCN)~\cite{chen2018incorporating} leverages stock relation to construct an isomorphic graph to aggregate neighbor stocks' feature through convolution operation in spectral domain. By introducing the attention mechanism, Graph Attention Network (GAT)~\cite{feng2019temporal} learns the importance of neighbor stocks to allocate attention properly. 

\begin{figure}[t]  
 \center{\includegraphics[width=9cm]  {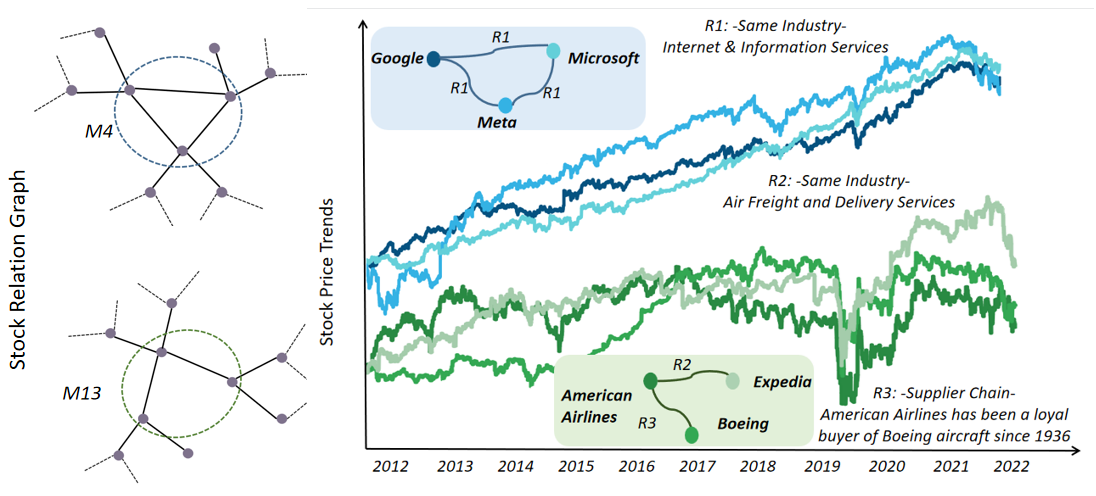}}
 \caption{Stocks within motif show correlated price trends in the last decade: Google, Microsoft and Meta are related under triangular motif M4. American Airlines, Expedia and Boeing are related under motif M13. }
 \label{relation2}
 \end{figure}

GNN-based methods typically address stock prediction as either a classification (on price movement direction) or a regression (on price value) task, rather than selecting the stocks with the maximum expected profit. 
Some papers~\cite{feng2019temporal, sawhney2021stock, ma2022fuzzy} indicate that a better regression result (low mean squared error) suggests a less profitable stock. Therefore, based on GNNs, some state-of-the-art (SOTA) 
methods~\cite{feng2019temporal, sawhney2021stock, ma2022fuzzy} jointly consider both predictive performance and the expected profit and directly optimized. 

However, the above SOTA methods for stock selection have two significant drawbacks. First, they only consider simple pairwise relation between two stocks.
However, in many real-world graphs (e.g., social networks), the minimal and functional structural unit of a graph is not a simple edge but a small network subgraph that involves more than two nodes \cite{benson2015tensor, ge2021mixed}, which is called higher-order structures (a.k.a motifs). There is substantial evidence that higher-order structures (a.k.a motifs) that are small subgraph connectivity patterns among several nodes (e.g., triangles, $4$-vertex cliques), are essential to understand the behaviour of many complex systems modelled by graphs~\cite{milo2002network, yaverouglu2014revealing, benson2016higher}. 
The higher-order structure (HoS) reveal functional property of a complex network and has been used in the study of biological protein~\cite{moore2008protein}, social networks~\cite{zhao2019ranking} and wireless communications~\cite{jiang2013review}. The stock graph consisting of higher-order structures have correlated price trends. As shown in Fig.~\ref{relation2},  Google, Microsoft and Meta are within a fully connected triangular motif \textit{M4} and share highly coincident price movements for the past decade. Expedia and Boeing are not directly connected, but their influence is indirectly connected via American Airlines, forming a sort of 'mutual friend' third-order motif \textit{M13}. 

The second drawback is that the majority of existing work of stock selection only focus on technical analysis (TA) and do not incorporate fundamental analysis (FA) \cite{luckieta2020fundamental}. In reality, both TA and FA are important. TA is primarily focused on analyzing historical price and volume data of stock to identify price patterns and trends, and often have shorter time horizons \cite{nti2020systematic}. On the other hand, FA aims to determine the intrinsic value of a stock by analysing financial health of this company from the perspective of long term \cite{xu2019stock}. For example, some key financial metrics are widely used to analyse company health from financial statements ~\cite{OU1989295}, such as profitability, growth, leverage, earning per share. These factors have been widely used and identified in financial domain to analyse stock price \cite{djazuli2017relevance}\cite{hull1999leverage}.

Motivated by them, in this paper, we propose a \underline{h}igher-order \underline{g}raph \underline{at}tention network with joint analysis for stock selection (H-GAT). We collected quarterly financial statements for over 9,000 stocks on the NYSE and Nsadaq over the past decade. H-GAT jointly considers both TA and FA to capture the sequential dependencies, meanwhile extract higher-order structures of stock graphs to learn stock embeddings. Specifically, we first feed both historical price (e.g., moving average of closed price and essential factors (e.g., leverage) from financial statement to a Long Short-Term Memory (LSTM) to learn a stock sequential embedding. Simultaneously, we represent higher-order stock relation with  GAT from a stock heterogeneous graph. Finally, we feed the concatenation of sequential embeddings and relational embeddings to a fully connected layer to obtain the ranking score of stocks. Our contributions are summarised as follows:

\begin{enumerate}
\item We propose to jointly consider both TA and FA for a stock ranking model.
\item We propose H-GAT to learn higher-order structures representation from stock heterogeneous graphs.
\item We conduct experiments on mean reciprocal rank, internal rate of return and Sharpe ratio to show the superior performance over state-of-the-art methods.
\end{enumerate}




 
\section{Related Work}

\subsection{Notations}
We denote scalars by lowercase letters, e.g., $d$, 
vectors by lowercase boldface letters, e.g., $\mathbf{d}$, matrices by  uppercase  boldface, e.g., $\mathbf{D}$. 
Let $G = (V, E)$ be an undirected unweighted graph (network) with $V$ = \{v$_1$, v$_2$, \dots, v$_n$\} being the set of $n$ nodes, i.e., $n$ = $|V|$, and $E$ = \{e$_1$, e$_2$, \dots, e$_n$\} being the set of edges connecting two nodes.

\subsection{Higher-order Structures} 

\begin{figure}[t]
 \center{\includegraphics[width=8.5cm] {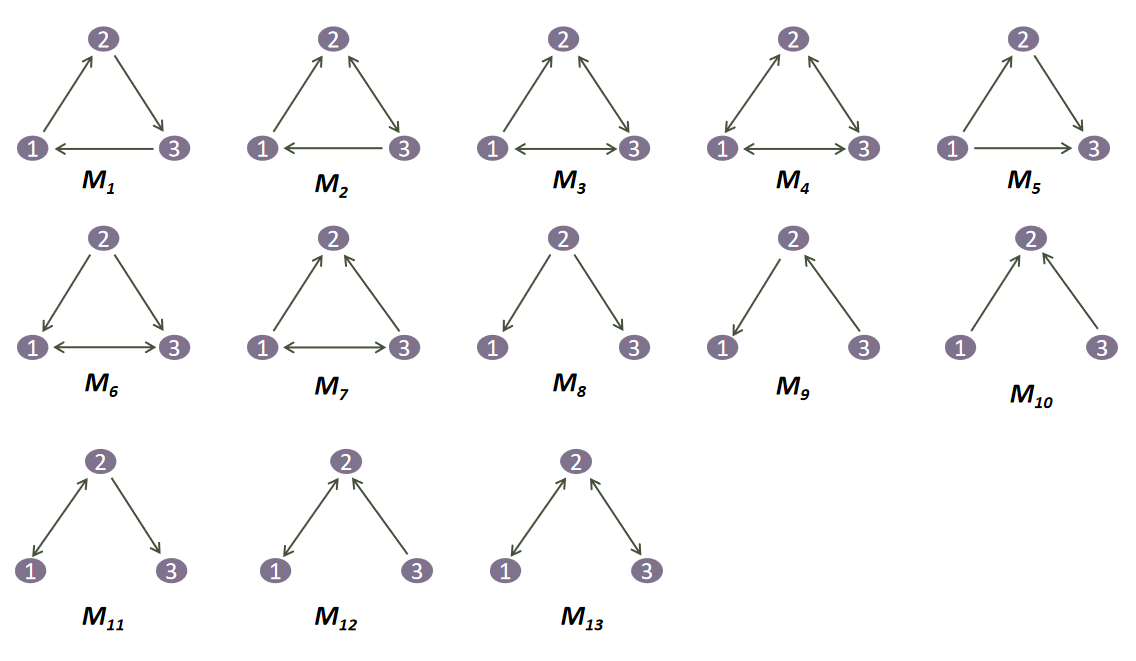}}  
 \caption{13 third-order structures.} \label{fig:motif}
 \label{motif}
 \end{figure}

 Graph analysis in terms of the edge structure connecting two nodes is natural since the edge is assumed to be the simplest unit structure in graphs. However, there is substantial evidence that higher-order structures (Fig. \ref{fig:motif}) that are small subgraph connectivity patterns among several nodes, are essential to understand the behaviour of many complex systems modelled by graphs 
 \cite{benson2016higher, milo2002network, ge2020unifying}.
Higher-order structures have been applied in the fields of protein classification and integration~\cite{yeger2004network}, neuron activity prediction~\cite{varshney2011structural}, and recommendations~\cite{gupta2014real, bai2019joint}. More importantly, higher-order structures allow for more flexible modelling. For instance, considering directions of edges, there exist
$13$ different {third-order structures},
but only two different second-order structures. Thus, the application can drive which
third-order structures to be preserved. 
The paper ~\cite{yu2022motifexplainer} points out that most existing GNN methods can identify important edges and nodes, but do not consider subgraph structures. They stipulate extraction rules to identify recurring important patterns in graphs. William \etal~\cite{underwood2020motif} defines the motif adjacency matrix on weighted directed networks and confirms the effectiveness of motif for spectral clustering algorithms.

\begin{myDef}
(Higher-order Structures). Suppose the directed weighted graph is made up of triples G= ($\boldsymbol\nu$,$\boldsymbol\varepsilon$,$\boldsymbol\omega$) where $\boldsymbol\nu$ stands for vertices, $\boldsymbol\varepsilon$ for edges, and $\boldsymbol\omega$ for edge weights. In the graph G, if a subgraph $G_{1}$= ($\boldsymbol\nu_{1}$,$\boldsymbol\varepsilon_{1}$) satisfies that $\boldsymbol\nu_{1} \subseteq  \boldsymbol\nu$ and $\boldsymbol\varepsilon_{1} \subseteq  \boldsymbol\varepsilon$. The subgraph $G_{1}$ can be regarded as a higher-order structure of G. Usually $\left \| \boldsymbol\nu\right \|$= 3 or 4, we mainly focus on 3-node network motifs.
\end{myDef}

\subsection{Fundamental Analysis and Technique Analysis}

Technical analysis (TA)\cite{schabacker2021technical} focuses on past and present market behaviour through charts or records in order to predict future price trends. It is based on the main content of the technical indicators calculated from the stock price, volume or rise and fall stock index. Most existing works only leverage TA, they collect stock price and trading volume data, feed into time series\cite{rounaghi2016investigation} \cite{herwartz2017stock}, machine learning\cite{patel2015predicting}\cite{shen2012stock} or deep learning\cite{singh2017stock}\cite{long2019deep}\cite{lin2018modelling} models to predict stock returns. 

However, TA only concerns the changes of the securities market itself and ignores the economic, political and other external factors\cite{petrusheva2016comparative}. Fundamental analysis (FA)\cite{luckieta2020fundamental} \cite{hull1999leverage} method evaluates general economic situation, the operation and management status of each company to study the value of stocks and measure the level of stock prices. Therefore, we believe that fundamental analysis is also indispensable. We aim to build a model like a real fund manager, which can not only grasp the specific buying time through technical analysis, but also analyze the intrinsic value of stocks and describe long-term upside space through fundamental analysis\cite{bonga2015need}.

\subsection{Graph Neural Networks}
Graph neural networks~\cite{9395439} abstract non-Euclidean data such as stock relationships into nodes and edges. Graph neural networks have been surprisingly successful in the fields of biological structures~\cite{fout2017protein} and social network recommendations~\cite{fan2019graph}. Mainstream graph neural networks include Graph Convolution Networks (GCN)~\cite{kipf2016semi}, Graph Attention Networks (GAT)~\cite{velivckovic2017graph} and Hypergraph~\cite{feng2019hypergraph}. Jiexia~\cite{ye2021multi} combines graph convolutional networks (GCN) and gated recursive units (GRUs) to propose a Multi-GCGRU framework. Multi-GCGRU extracts the cross effect of time series by GRU and stock interaction by GCN, which proved effective in the Chinese stock market. Fuli Feng~\cite{feng2019temporal} has successfully modelled stock relationships and stock prices as time-sensitive graphs and proposed a TGC component to rank stocks. However, the stock relation source of TGC is not aggregated, that is, the model can only select the stock relation from either wikidata or industry. Raehyun Kim~\cite{kim2019hats} divided stock relationships into different sources and proposed hierarchical attention networks (HATS). Kun~\cite{huang2022ml} proposes a multilevel graph attention network (ML-GAT) to select and encode different relationships to build isomorphic stock graphs. Up to now, most of the graph neural network models for stock graph construction still stay in the ordinary graph. Ramit Sawhney~\cite{sawhney2021stock} proposed that ordinary graphs linking stocks in pairs might have information loss. The high-order relationship between stocks is not fully explored.

\section{Methodology}

 \begin{figure*} [t] 
 \center{\includegraphics[width=18cm]  {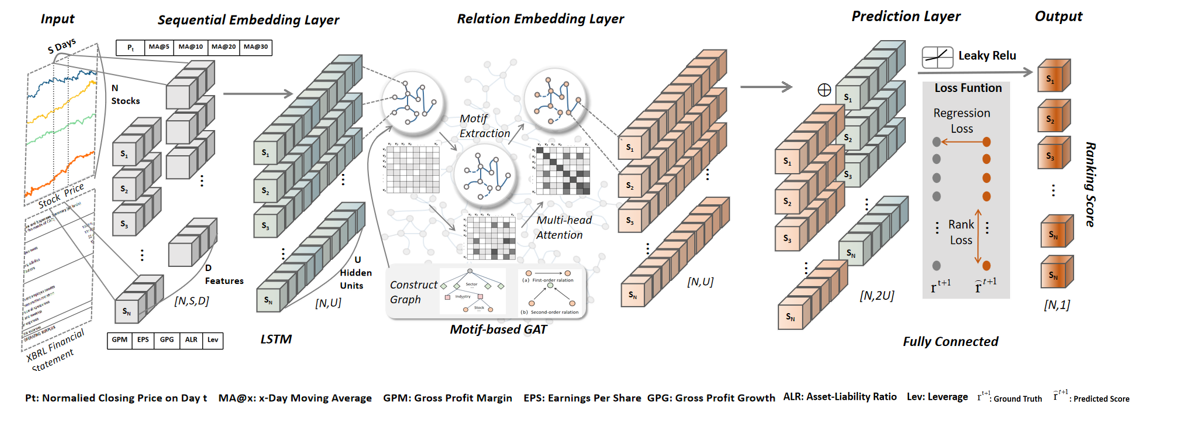}}  
 \caption{Framework of Higher-order Graph Attention Network (H-GAT). H-GAT contains three layers, the sequential embedding layer, the relational embedding layer (the pair relationship and the high order model relationship) and the prediction layer, which outputs the next day's stock investment return.}
 \label{framework}
 \end{figure*}

 \subsection{Framework}
 
Our H-GAT aims to train a stock ranking model, which consists of the sequential embedding layer, relational embedding layer and prediction layer. We first collect fundamental factors from financial statements and technical factors. In the sequential layer, given two types of factors of $n$ stocks among $t$ time period, the LSTM learns the sequential embedding for each stock.
 To depict the stock relation, building on the work \cite{feng2019temporal}, we extend stock relation types. We construct a higher-order adjacency matrix $\mathbf{A}_h$ as the input to the GAT model to generate relational embedding. The relational data set includes C kinds of industry relations and wiki relations. Finally, we feed the concatenation of sequential embeddings and relational embeddings to a fully connected layer to obtain the ranking score of stocks. An illustration of the framework H-GAT can be found in Fig.\ref{framework}.


\begin{table}[t]
\renewcommand{\arraystretch}{1}
\centering
\begin{tabular}{p{2.5cm}|p{5cm}}
\hline
\textbf{Notation}& \textbf{Description}\\
\hline

$\mathbf{\chi ^{t}} \epsilon \mathbb{R}^{N\times S\times D}$ & sequential input on trading day t \\
$\mathbf{E_{s}^{t}} \in \mathbb {R}^{N\times U}$& sequential embedding output\\
$\mathbf{E_{r}^{t}} \in \mathbb {R}^{N\times U}$& relation embedding output\\
$\mathbf{A_{wiki}} \in \mathbb {R}^{N\times N}$& adjacency matrix from C channels\\
$\mathbf{\tilde{\boldsymbol\zeta}_{M,A}}$ & anchored automorphism sets\\
$\mathbf{A_{MAM}}\in \mathbb {R}^{N\times N}$ & motif adjacency matrix\\
$\mathbf{r^{t+1}}$,$\mathbf{\hat{r}^{t+1}}\in \mathbb{R}^{N}$&ground truth and predicted return ratio\\
$\mathbf{\boldsymbol\omega},b$ & weights and bias to be learned\\

\hline
\end{tabular}
\label{notations}
\caption{Key Notations}
\end{table}


\subsection{Sequence Embedding Layer}
The sequence embedding layer considers the strong temporal dynamics of stock market, and jointly leverage fundamental factors and technical factors. In this layer, we use the LSTM that is a type of recurrent neural network (RNN) architecture, which is specifically designed to address the limitations of traditional RNNs in learning long-range dependencies and avoiding the vanishing gradient problem.


 At each time-step $t$, $\mathbf{x}^{\mathrm{t}} \in \mathbb{R}^d$ denotes an input vector, where $d$ is the input dimension. The input vector $\mathbf{x}$ concatenates both fundamental factors and technical factors. We collect fundamental factors from quarterly financial statements, and the collection will be describe in Sec \ref{sec:fundamental}. For technical factors, we standardize 1-day return ratio and calculate moving averages (i.e., 5, 10, 20, and 30 days). 
 

In this paper,  we utilize vanilla LSTM~\cite{bao2017deep} that is composed of three gates and two states: input gate $\mathbf{i_{t}}$, output gate $\mathbf{o_{t}}$, forget gate $\mathbf{f_{t}}$, cell state $\mathbf{c_{t}}$ and hidden state $\mathbf{h_{t}}$. The LSTM takes in features generating from $[t-s,t]$ days historical sequence data, detailed mechanisms can be defined via the following equations:

\begin{equation}
    \mathbf{f_{t}} = \sigma (\mathbf{W_{f}}\cdot [\mathbf{h_{t-1}},\mathbf{x_{t}}]+\mathbf{b_{f}})
\end{equation}
\begin{equation}
    \mathbf{i_{t}} =\sigma (\mathbf{W_{i}}\cdot [\mathbf{h_{t-1}},\mathbf{x_{t}}]+\mathbf{b_{i}})
    \end{equation}
\begin{equation}
    \mathbf{\widetilde{c_{t}}} = tanh(\mathbf{W_{c}}\cdot [\mathbf{h_{t-1}},\mathbf{x_{t}}]+\mathbf{b_{c}})
    \end{equation}
\begin{equation}
    \mathbf{c_{t}} =\mathbf{f_{t}}\ast \mathbf{c_{t-1}}+\mathbf{i_{t}}\ast \mathbf{\tilde{c_{t}}}
    \end{equation}
\begin{equation}
    \mathbf{o_{t}} =\sigma (\mathbf{W_{o}}[\mathbf{h_{t-1}},\mathbf{x_{t}}]+\mathbf{b_{o}})
    \end{equation}
\begin{equation}
    \mathbf{h_{t}}=\mathbf{o_{t}} \ast tanh(\mathbf{c_{t}})
\end{equation}
where $\mathbf{W_{f}},\mathbf{W_{i}},\mathbf{W_{c}},\mathbf{W_{o}} \in \mathbb {R}^{u\times u}$ are mapping matrices and U is the number of hidden units,
$\mathbf{b_{f}},\mathbf{b_{i}},\mathbf{b_{c}},\mathbf{b_{o}}\in \mathbb {R}^{u}$ are bias vectors. 

We use LSTM  to processes the input $\mathbf{x_{t}}$ and former hidden state $\mathbf{h_{t-1}}$. The forget gate $\mathbf{f_{t}}$ and the input gate $\mathbf{i_{t}}$ decide if the information need to forget or remain, while $\mathbf{\widetilde{c_{t}}}$ represents candidate information to be injected into the cell state $\mathbf{c_{t}}$. The output gate $\mathbf{o_{t}}$ determines the output and is used to update the next hidden state $\mathbf{h_{t}}$. Benefiting from its memory states, LSTM is proficient at capturing long-term dependence and handling the prediction of multi-variable time series. We obtain the output of the hidden state $\mathbf{h_{i}^{t}}$ as sequential embedding $\mathbf{e_{i}^{t}} \in \mathbf{E_s^{t}}^{n \times u} = [\mathbf{e_{1}^{t}}, \mathbf{e_{2}^{t}}, \cdots, \mathbf{e_{n}^{t}}]$, where $n$ is the number of stocks and $u$ is the dimension of hidden units. Thus, we have:
\begin{equation}
    \mathbf{E_s^{t}} = \text{LSTM}(\mathbf{X}^{t})
\end{equation}

\subsection{Higher-order Graph Embedding Layer}

The HoS are fundamental to gain insights into stock graphs. Additionally, the usage of the HoS allows for greater modelling flexibility. Despite its key importance, the HoS have not been well integrated into the existing works. Motivated by it, our higher-order graph embedding layer aims to capture the HoS that are used to learn node embeddings with a graph attention network. 

Specifically, this layer incorporates the relational adjacency matrix $\mathbf{A}$ and the sequence embedding $\mathbf{E_{s}^{t}}$, and outputs the relational sequential embedding as $\mathbf{E_{r}^{t}}$. Building on the paper \cite{feng2018enhancing}, we significantly extend relation industry and Wiki types to build stock graphs. The collection process will be detailed in Section \ref{data collection}. We use the multi-hot binary adjacency matrix $\mathcal{A}_w \in \mathbb{R}^{n \times n \times c_1}$ and $\mathcal{A}_i \in \mathbb{R}^{n \times n \times c_2}$ to encode the Wiki relation and industry relation respectively, where $c_1$ and $c_2$ are the number of Wiki relation and industry relation. We compress multi-hot binary
adjacency matrix  $\mathcal{A}_w \in \mathbb{R}^{n \times n \times c_1}$ via a fully connected (FC) layer as follows:

\begin{equation}
    \mathbf{A_{w}}^{n\times n}=f_{c1}(\mathcal{A}_w).
\end{equation}


$\mathbf{A_{w}}$ construct $G_{wiki}$=($\boldsymbol\nu_{M}$,$\boldsymbol\varepsilon_{M}$), which is used to construct a motif adjacency matrix $\mathbf{A_{h}}$.
We first define 0-1 edge indicator matrices $\mathbf{J_{ij}}$ in directed weighted graphs for single-edge $\mathbf{J_{s}}$, double-edge $\mathbf{J_{d}}$, missing-edge $\mathbf{J_{0}}$ and vertex-distinct $\mathbf{J_{n}}$. $\mathbb{I}$ is indicator function:

\begin{equation}
\begin{split}
\mathbf{J_{ij}} 
& =\mathbb{I}\{(i,j)\ \in \ \boldsymbol\varepsilon \}\\
\mathbf{(J_{s})_{ij}} 
& =\mathbb{I}\{(i,j)\ \in \ \boldsymbol\varepsilon\ and\ (j,i)\ \notin \ \boldsymbol\varepsilon \}\\
\mathbf{(J_{d})_{ij}} 
& =\mathbb{I}\{(i,j)\ \in \ \boldsymbol\varepsilon\ and\ (j,i)\ \in \ \boldsymbol\varepsilon \}\\
\mathbf{(J_{0})_{ij}} 
& =\mathbb{I}\{(i,j)\ \notin\ \boldsymbol\varepsilon\ and\ (j,i)\ \notin\ \boldsymbol\varepsilon\ and\ i\neq\ j\}\\
\mathbf{(J_{n})_{ij}} 
& =\mathbb{I}\{i\neq j\}\\
\end{split}
\end{equation}
Then, based on edge indication matrix $\mathbf{J_{ij}}$, we update the weight adjacency matrix $\mathbf{W_{ij}}$ for directed edges $\mathbf{W}$, single edges $\mathbf{W_{s}}$, and double edges $\mathbf{W_{d}}$, respectively:

\begin{equation}
    \begin{split}
        \mathbf{W_{ij}}
        & =\boldsymbol\omega((i,j))\mathbb{I}\{(i,j)\ \in \boldsymbol\varepsilon \}\\
\mathbf{(W_{s})_{ij}}
& =\boldsymbol\omega((i,j))\mathbb{I}\{(i,j)\ \in \boldsymbol\varepsilon\ and\ (j,i)\ \notin \ \boldsymbol\varepsilon\}\\
\mathbf{(W_{d})_{ij}}
& =(\boldsymbol\omega((i,j))+\boldsymbol\omega((j,i)))\ \mathbb{I}\{(i,j)\in \boldsymbol\varepsilon\\
& \ and\ (j,i)\in \boldsymbol\varepsilon\}\\
    \end{split}
\end{equation}

Higher-order structures in $G_{wiki}$=($\boldsymbol\nu_{M}$,$\boldsymbol\varepsilon_{M}$) including empty $e^{0}$, unilateral $e^{s}$ and bilateral $e^{d}$ are expressed as follows ( $k_{1}\leq k_{u}<k_{v}\leq k_{m}$):
\begin{equation}
 \begin{split}
e^{0}
&=\{(k_{u},k_{v}): (k_{u},k_{v})\notin \boldsymbol\varepsilon_{M},(k_{v},k_{u}) \notin \boldsymbol\varepsilon_{M}\}\\
e^{s}
&=\{(k_{u},k_{v}): (k_{u},k_{v})\in \boldsymbol\varepsilon_{M},(k_{v},k_{u}) \notin \boldsymbol\varepsilon_{M}\}\\
e^{d}
&=\{(k_{u},k_{v}): (k_{u},k_{v})\in \boldsymbol\varepsilon_{M},(k_{v},k_{u}) \in \boldsymbol\varepsilon_{M}\}\\
 \end{split}
\end{equation}

When we match motif graph $G_{m}$ with $\boldsymbol\nu_{M}=(k_{1},...,k_{m})$ in graph G with $\boldsymbol\nu=(1,...,n)$, if we set any 2 node pairs such as $i=k_{1},j=k_{m} (i,j \in \boldsymbol\nu)$, we need to search all mappings of  $u\rightarrow k_{\sigma u}$. Indicator matrix \textbf{J} and weighted adjacency matrix \textbf{W} are calculated as follows:
\begin{equation}
    \mathbf{J_{k,\sigma}}=\prod_{e^{0}}(\mathbf{J_{0}})_{k_{\sigma u},k_{\sigma v}}\prod_{e^{s}}(\mathbf{J_{s}})_{k_{\sigma u},k_{\sigma v}}\prod_{e^{d}}(\mathbf{J_{d}})_{k_{\sigma u},k_{\sigma v}}
\end{equation}
\begin{equation}
    \mathbf{W_{k,\sigma}}=\sum_{e^{s}}(\mathbf{W_{s}})_{k_{\sigma u}k_{\sigma v}}+\sum_{e^{d}}(\mathbf{W_{d}})_{k_{\sigma u}k_{\sigma v}}
\end{equation}

\begin{myDef}
 (Anchored automorphism sets). Suppose that the motif is $G_{m}$=($\boldsymbol\nu_{M}$,$\boldsymbol\varepsilon_{M}$) and $\mathbf{\boldsymbol\zeta_{M}}$ is the permutations of vertices in $\boldsymbol\nu_{M}=\{k_{1},...,k_{m}\}$. We define the anchored permutations as $\mathbf{\boldsymbol\zeta_{M,A}}=\{\sigma \in \mathbf{\boldsymbol\zeta_{M}}:\{k_{1},k_{m}\}\subseteq \sigma (A)\}$. Then the equivalent conditions $\sim$ defined on $\mathbf{\boldsymbol\zeta_{M,A}}$ represents for automorphism of $G_{m}$. The quotient set equivalent to $\mathbf{\boldsymbol\zeta_{M,A}}$ is anchored automorphism sets $\mathbf{\tilde{\boldsymbol\zeta}_{M,A}}$: $\mathbf{\tilde{\boldsymbol\zeta}_{M,A}}$=$\mathbf{\boldsymbol\zeta_{M,A}}/\sim$.
\end{myDef}
Then, for anchored node pairs $i=k_{1},j=k_{m} (i,j \in \boldsymbol\nu)$, the higher-order adjacency matrix $\mathbf{A_{h}}$ in the graph $G_{wiki}$ is:

\begin{equation}
     \mathbf{A_{h}}=\frac{1}{\left | \boldsymbol\varepsilon \right |}\sum_{\sigma\in \mathbf{{\tilde{\boldsymbol\zeta}_{M,A}}} }\sum_{\{k_{2}...k_{m-1}\}\subseteq \boldsymbol\nu}J_{k,\sigma }W_{k,\sigma},
\end{equation}
We use anchored automorphism sets to guarantee recognizing the $m$-order structures correctly and avoid duplicate count for the same vertex set by constraining $\sigma\in\mathbf{{\tilde{\boldsymbol\zeta}_{M,A}}}$. We then construct a higher-order matrix $\mathbf{A}^{(i)}_h$ for the type $i$ from the matrix operation of $\mathbf{A_{w}}$. The tensor $\mathcal{A}_h$ consists of all higher-order matrices from 1 to $k$ where $k$ is the total higher-order structure types.

We then concatenate $\mathbf{A_{h}}$ with $\mathbf{A_{w}}$ and $\mathbf{A_{i}}$, and generate final relation adjacency matrix $\textbf{A}$ through different FC layers:

\begin{equation}
    \mathbf{A}=f_{c2}(\mathcal{A}_w\bigoplus \mathcal{A}_i\bigoplus \mathcal{A}_h)
\end{equation}

Then for attention coefficient, instead of calculating dot product similarity or cosine similarity\cite{luo2018cosine}, we propose a custom function $g(a_{ji},e_{i}^{t},e_{j}^{t})$. Feeding in sequential embedding $\mathbf{E_{s}^{t}}$ and adjacency matrix $\mathbf{A}$, $g(a_{ji},e_{i}^{t},e_{j}^{t})$ can implicitly modelling relation strength:

\begin{equation}
    g(\mathbf{a_{ji}},\mathbf{e_{i}^{t}},\mathbf{e_{j}^{t}})=\phi (\mathbf{w^{T}}[\mathbf{e_{i}^{t^{T}}},\mathbf{e_{j}^{t^{T}}},\mathbf{a_{ij}^{T}}]^{T}+b)
\end{equation}
where $\phi$ is softmax activation function, $\mathbf{a_{ij}}$ is relation vector and $\mathbf{e_{i}^{t}},\mathbf{e_{j}^{t}}$ represent for sequential embeddings for stock i and j at day t.

To fully mine information in different subspaces and stabilise results, we compute multi-head attention. We then use sigmoid function $\sigma$ to normalize the output of $g(\mathbf{a_{ji}},\mathbf{e_{i}^{t}},\mathbf{e_{j}^{t}})$ and add more non-linearities. K-head graph attention network's embedding propagation is as follows:


\begin{equation}
    \mathbf{\bar{e_{i}^{t}}}=\sum_{{j|sum(a_{ij})>0}}\frac{\sigma(\frac{1}{K}\sum_{k-1}^{K}g^{k}(\mathbf{a_{ji}},\mathbf{e_{i}^{t}},\mathbf{e_{j}^{t}}))}{d^{k}_{j}}\mathbf{e_{j}^{t}},
\label{multi-attention}
\end{equation}
where $g^{k}(\mathbf{a_{ji}},\mathbf{e_{i}^{t}},\mathbf{e_{j}^{t}})$ is the $k$-th attention mechanism with a parameter $\mathbf{w}_k$. We can then use this way to generate higher-order graph embedding $\mathbf{E_{r}^{t}} \in \{\bar{\mathbf{e}}_1^t, \bar{\mathbf{e}}_2^t, \cdots, \bar{\mathbf{e}}_n^t\}$.



\subsection{Prediction layer}
We feed both sequential embedding $\mathbf{E_{s}^{t}}$ and relational embedding $\mathbf{E_{r}^{t}}$ to 
predict stock return rate of stocks for the next day $\hat{r}^{t+1}$ through a full connected layer $f_{c3}$:
\begin{equation}
   \mathbf{\hat{r}^{t+1}} = f_{c3}(\mathbf{E_{s}^{t}},\mathbf{E_{r}^{t}}),
\end{equation}
where $\mathbf{\hat{r}^{t+1}} =\left[r_1^{t+1}, \cdots r_n^{t+1}\right]$ is the predicted return ratio of the stock from 1 to $n$ for the $t+1$ day. 
Following the paper~\cite{sawhney2021stock}, the loss function includes regression loss and paired ranking loss to conduct model optimizing towards correct ranking position. The regression loss are used to anchor ground truth, meanwhile the paired ranking loss promotes maintaining the same relative order between the predicted ranking scores of a stock pair as in the ground-truth. The weight between regression loss and pairwise ranking loss is balanced by hyperparameter $\alpha$:
\begin{equation}
\begin{split}
    \mathcal{L}(\mathbf{\hat{r}^{t+1}},\mathbf{r^{t+1}}) = 
    &\mathcal{L}_{reg}(\mathbf{\hat{r}^{t+1}},\mathbf{r^{t+1}})+\mathcal{L}_{rank}(\mathbf{\hat{r}^{t+1}},\mathbf{r^{t+1}})\\=
    & \left \| \mathbf{\hat{r}^{t+1}}- \mathbf{r^{t+1}}\right \|^{2}+\alpha \sum_{i=0}^{n}\sum_{J=0}^{n} max\\
    & (0,-(\hat{r}_{i}^{t+1}
    -\hat{r}_{j}^{t+1})(r_{i}^{t+1}-r_{j}^{t+1})),
    \end{split}
\end{equation}
where $\max({0, a})$ represents the maximum value either 0 or $a$.
 
\section{Data Collection}\label{data collection}
To capture spatial and spatio property of stock data, the existing stock dataset includes sequential data and relational data. However, 
most existing datasets only have a short time span and do not include factors for FA \cite{agrawal2013state,feng2019temporal,feng2018enhancing }. Therefore, in this paper we construct a comprehensive stock dataset with long term span over one decade and five significant factors from financial statements for FA. The benefits of our datasets are shown below: 
\begin{itemize}
\item Factors for FA: In addition to sequential data such as stock price, non-euclidean relation data and financial statement data are also collected.
\item Long-time span: The data set contains stock prices for over a decade, ranging from January 3, 2012 to July 5, 2022.
\item Extensive stock coverage: We collect all stocks traded on NYSE and NASDAQ, the two largest stock markets by market capitalization in the world, with 3,404 and 5,717 stocks, respectively.
\end{itemize}

\subsection{Sequential Data} 
\subsubsection{Stock Price}

Our constructed data is based on the daily historical closed price from January 3, 2012 to July 5, 2022. The ground truth is set as one-day return ratio $r_i^{t+1}=\left(p_i^{t+1}-p_i^t\right) / p_i^t$, where $p_i^t$ is the closing price at the day $t$.  We normalize the price of each stock by dividing it by its highest value over the entire dataset's time period. Besides the normalized closing price, we compute four additional sequential features: moving averages for 5, 10, 20, and 30 days, representing weekly and monthly trends. 

We collect data of 5717 and 3404 stocks on the NASDAQ and NYSE markets. Then we select the stocks based on three criteria: 1) trading records are recorded on 98\% or more of the trading days; 2) never trading below \$5, eliminating risky penny stocks; 3) no missing quarterly financial reports. After selecting, the number of stocks is 328 on NASDAQ and 455 on NYSE. Then we chronologically divide the datasets into train, validation and test sets.
After that, we split the whole dataset into the train, validation and test set. The statistics about stock price data is shown in the the upper part of the Table \ref{tab:stats}. In addition, to guarantee the reliability and reusability of the dataset, we collect data by using official APIs rather than third-party APIs. Google has retained the interface to Google Finance as a function within Google Sheet. We use Google Sheet API~\footnote{\url{https://developers.google.com/sheets/api/guides/concepts}} to write, update and download stock trading price from Google Sheet.


\subsubsection{FA Factors in Financial Statements} \label{sec:fundamental}

Financial statements used to be disclosed on paper or photocopied as PDF. In 2009, the U.S. Securities and Exchange Commission (SEC) announced official start of the XBRL reporting format ~\cite{abdullah2009institutionalisation}. This is a revolution in the securities industry, where previously unstructured financial statements such as plain text or pictures could be searched and checked, and then spreadsheets can be exported through XBRL for analysis. The three main financial statements are the balance sheet, income statement and cash flow statement\cite{subramanyam2014financial}. We extract characteristics in Table \ref{financial features} from the following aspects to reflect a company's financial status:

Profitability is the most important indicator for investors and directly reflects the level of industry and product rivalry. Gross profit margin usually indicates whether the company's products have a certain brand premium or lower direct production costs~\cite{nariswari2020profit} ~\cite{mahdi2020influence}. Earnings per share is the income earned per common share ~\cite{patell1976corporate}. These two indicators reflect company's profitability from the perspective of production and operation and profit distribution.

 Growth analysis focuses on the time when a company is expanding. Many small businesses struggle with profitability in the beginning stages of growth, but due to high market potential and an ability to seize chances, they have a larger chance of developing later and earning more money.

 Debt risk analysis is used to assess a company's financial security. The asset-liability ratio is positively correlated with the long-term solvency of enterprises~\cite{holz2002impact}. Appropriate leverage can boost corporate earnings, but too much leverage puts the company at risk of debt default and insolvency. One of the major causes of the 2008 financial crisis was the excessive accumulation of financial leverage in the banking system~\cite{shahzad2015financial}.

\renewcommand{\arraystretch}{1.2}\label{tab:stats}
\begin{table}[t] 
\centering
\begin{tabular}{m{0.07\textwidth}|m{0.035\textwidth}<{\centering}|m{0.075\textwidth}<{\centering}|m{0.075\textwidth}<{\centering}|m{0.075\textwidth}<{\centering}}
\hline
\hline
\multicolumn{5}{c}{\textbf{Sequential Data}}\\
\hline
\hline
 &Stocks&Train 01/03/2013 12/31/2015&Validation 01/04/2016 12/31/2016&Test 01/03/2017 12/31/2017\\
\hline
\textbf{NASDAQ} &328& 756&252&237\\
\hline
\textbf{NYSE} &455 & 756&252&237\\
\hline
\hline
\multicolumn{5}{c}{\textbf{Relation Data}  $_{*Industry+Wiki+Motif}$}\\
\hline
\hline
 &Nodes&Edges &\multicolumn{2}{c}{Relation Type} \\
\hline
\textbf{NASDAQ} &328&\num{2.38e4}&\multicolumn{2}{c}{$169_{*(84+80+5)}$}\\
\hline
\textbf{NYSE} &455 &\num{5.29e4}&\multicolumn{2}{c}{$206_{*(101+99+6)}$}\\
\hline
\end{tabular}
\caption{Statistics of our constructed sequential and relational datasets. Our datasets includes five significant factors for FA, long-time span and extensive stock coverage.}
\label{tab}
\end{table}


\subsubsection{Sector-Industry Relation}

 Morgan Stanley Capital International (MSCI) and Standard \& Poor's jointly defined Global Industry Classification Standard (GICS) in 1999 to establish a global standard for classifying listed companies~\cite{barra2009global}. GICS classifies the financial securities industry into four levels: 11 sectors, 24 industry groups, 69 industries, and 158 sub-industries~\cite{chan2007industry}. Companies in the same industry tend to have similar stock price performance because they are under comparable customer environment and market risk. Full GICS stock industry classification can be downloaded at NASDAQ \footnote{\url{http://www.nasdaq.com/screening/companies-by-industry.aspx}}. Based on GICS industry classification, we extract 84 and 101 relationships from the NASDAQ and NYSE stock market, respectively. 
 
 We calculate the coverage ratio of stock pairs with at least one relation connection. Industry relations are relatively sparse in both NASDAQ and NYSE stock market, 14.89\% and 16.78\% of stock pairs are connected through industry relationships, respectively.

\begin{table}[t]
\renewcommand{\arraystretch}{2}
\centering
\begin{tabular}{c|c|c}
\hline
&\textbf{Features}& \textbf{Equation}\\
\hline
\multirow{2}*{Profitability}& Gross Profit Margin&$$\scalemath{0.9}{\frac{GP_t}{R_t} }$$   \\
\cline{2-3}
& Earnings Per Share&$$\scalemath{0.9}{\frac{NI_t - PSD_t}{AOS_t}}$$  \\
\hline
Growth& Gross Profit Growth&$$\scalemath{0.9}{\frac{GP_t - GP_{t-1}}{GP_{t-1}}}$$ \\
\hline
\multirow{2}*{Debit Risk}& Asset-Liability Ratio& $$\scalemath{0.9}{\frac{AS_t}{L_t}}$$\\
\cline{2-3}
& Leverage& $$\scalemath{0.9}{\frac{AS_t}{E_t}}$$  \\
\hline
\end{tabular}
\label{Financial Statement Metric}
\caption{Financial Statement Metrics. We extracted five new features from the earnings report, taking into account corporate earnings, growth and debt risk.}
\label{financial features}
\end{table}

\subsubsection{Wiki Relation}

\renewcommand{\arraystretch}{1.2}
\begin{table}[t]
\centering
\begin{tabular}{m{0.07\textwidth}|p{0.03\textwidth}<{\centering}|p{0.30\textwidth}}
\hline
Relation &Order & Description  \\
\hline
P127 & 1& owned by (P127):owner of the subject\\
\hline
P31\_P127 &2 & instance of (P31):that class of which this subject is a particular example and member.\\
\hline
P3320\_P112 &2 & board member (P3320):member(s) of the board for the organization
founded by (P112):founder or co-founder of this organization, religion or place \\
\hline
\end{tabular}
\caption{Wiki Relation Example}
\label{Wiki Relation Example}
\end{table}

As a rich source of entity relationships, Wikidata Stores over 99 million items and 10250 properties~\cite{vrandevcic2014wikidata}. We define first-order and second-order relation query: First-order relationships exist when two firms are directly connected. There is a strong correlation between companies that are directly connected, but the number of relationships is limited. Second-order relationships give a flexible way of relation source. Second-order relation refers to the fact that both entities are related to an item. 

SPARQLWrapper is Python's SPARQL endpoint interface, which can automatically query Wikidata and return JSON files. The initial query of NASDAQ and NYSE stock markets showed 867 and 228 stock relationships. We manually filter out valueless relationships, such as the same time zone, the same legal form of incorporation of the firm, or the number of followers on a social platform that matches the ID of another company. These relations contain negligible information and would be noisy to our model. After filtering, the NASDAQ and NYSE markets retain 80 and 99 stock relationships, respectively. 
Wiki relationships are dense compared to industry relationships. On the NASDAQ and NYSE stock market, the coverage ratio is 44.28\% and 51.11\%, respectively. The statistics of stock relation data is shown in the the upper part of the Table \ref{tab:stats}.

\section{Experiment}

 In this section, we conduct experiments to answer the following four research questions. 
 
 \begin{itemize}
 \item \textbf{RQ1}: What is the profitability ability of H-GAT compared with baselines?
  \item \textbf{RQ2}: Do the higher-order structures benefit to our H-GAT?
 \item \textbf{RQ3}: Do the factors of FA effectively enhance the performance of H-GAT?
  \item \textbf{RQ4}: Is H-GAT consistently profitable over long time span?
\end{itemize}

\subsection{Experimental Setting}
Following \cite{dixon2017classification}, our trading strategy involves buying and selling stocks based on their rankings. Specifically, we rank stocks at the end of each trading day $t$ within the tested date range (January 3, 2017, to December 31, 2017). The top 1, 5, and 10 ranked stocks are selected for trading. Traders buy these stocks at the closing price on day $t$ and sell them at the closing price on day $t+1$. Each transaction follows a simple interest investment model with a fixed principal of \$1,000, we use Internal Return
Ratio (IRR) for evaluation. Our trading experiments are based on the below assumptions related to real financial stock markets:

\begin{itemize}
    \item We assume that the stock market has high liquidity, enabling immediate settlement for all purchases and sales. Noting that real stock markets can experience reduced liquidity during bear markets or periods of weak market sentiment. In this paper, we ignore market liquidity and do not consider the impact of market efficiency on algorithm performance.
    \item We assume that there are no commission charges, despite engaging in high-frequency stock trading. Although commission charges can vary among brokers, they typically range around \$0.005 per share or \$5 per trade. High-frequency trading, characterized by high speeds and turnover rates, often incurs substantial commission charges. 
    \item We consider profits earned by investors as not reinvested, and the investment principal remains fixed. Consequently, all IRR calculations are based on simple interest rather than compound interest. In reality, investors can choose to reinvest their profits, potentially resulting in higher returns.
\end{itemize}

\subsection{Performance Metrics}
To evaluate our H-GAT thoroughly, we compare it with regression accuracy, ranking accuracy, profitability, and capacity for risk-return balancing. We describe 
 the four evaluation metrics below: 
\begin{itemize}
    \item Mean Square Error (MSE) : This is a commonly used evaluation metric of regression models, which is suitable for evaluating the gap between prediction value and ground truth:
    $ \text{MSE}=\frac{1}{|n|}\sum_{n}^{i=1}(r_{test}^{(i)}-\hat{r}_{test}^{(i)})$.
    \item Mean Reciprocal Rank (MRR) : 
     It measures the quality of the ranked list of search results by considering the rank of the first relevant result retrieved in response to a query. The MRR score ranges from 0 to 1, with higher values indicating better performance. A score of 1 indicates that the first relevant result is always retrieved as the top result, while a score of 0 indicates that the relevant result is never found in the top $k$ results. It is the equation
    $\text{MRR}=\frac{1}{n}\sum_{i=1}^{n}\frac{1}{k_{i}}$, where  $k_i$ is the rank of the first relevant result for query $i$.
    \item Internal Rate of Return (IRR) : It is the rate of return that an investment is expected to generate over its lifetime. It represents the average annual growth rate of the investment. IRR is the discount rate at which the cumulative net present value is zero. IRR is calculated when net present value (NPV) is zero: $\text{NPV}=\sum_{t=0}^{n}\frac{c_{t}}{(1+r)^{t}}$, where $c_t$ is the cash flow in the $t^{th}$ day, $r$ is the value of IRR. 
    \item Sharpe Ratio (SR) : This is a commonly used metric for fund performance evaluation. It refers to how much excess return (i.e., the portion of return ratio $R_{p}$ that exceeds the risk-free $R_{f}$ interest rate such as treasury bonds) can be generated per unit of risk increase: $\text{SR}=\frac{R_{p}-R_{f}}{\sigma _{p}}\label{sp}$.
  
\end{itemize}

\subsection{Hyperparameter Tuning}
Hyper-parameters are tuned via grid search selection. For all models, we use Adam optimizer\cite{kingma2017adam} and the same learning rate 0.001. 
 Each experiment is repeated five times, and the average performance is reported. The detailed hyperparameters are shown 
 in the Table \ref{tab:hyp}.
 
\begin{table}[t!] 
\centering 
\begin{tabular}{m{0.30\textwidth}|m{0.15\textwidth}<{\centering}}
\hline
\multicolumn{2}{c}{\textbf{Hyper-parameters Range}}\\
\hline
Length of historical sequence for feature l& [2,\ 4,\ 8,\ 16,\ 32]  \\
\hline
Number of hidden units u& [16,\ 32,\ 64,\ 128,\ 256] \\
\hline
Loss function's equilibrium coefficient $\alpha$ & [0.1,\ 1,\ 5,\ 10,\ 15,\ 20] \\
\hline
Number of heads in multi-head attention k & [1, 2, 3, 4, 5, 6, 7, 8] \\
\hline
\end{tabular}
\caption{Hyper-parameter settings.}\label{tab:hyp}
\label{tab}
\end{table}

\begin{table*}[h!]
\renewcommand{\arraystretch}{1.2}
\centering
\begin{tabular}{m{0.15\textwidth}|m{0.20\textwidth}<{\centering}|m{0.13\textwidth}<{\centering}|m{0.13\textwidth}<{\centering}}
\hline
\hline
\multirow{2}*{\ }& \multicolumn{3}{c}{\textbf{NASDAQ}}  \\
\cline{2-4}
& \textbf{MRR}& \textbf{IRR}& \textbf{SR}\\
\hline
Vanilla LSTM)&$3.64e^{-2}\pm 1.04e^{-3}$ &$0.13\pm0.62$&$ 0.56\pm 0.52$\\
\hline
SFM&$2.33e^{-2}\pm 1.07e^{-2}$ &$-0.25\pm0.52$&$ 0.23\pm 0.15$\\
\hline
GCN &$3.24e^{-2}\pm 3.21e^{-3}$ &$0.11\pm0.06$&$ 0.55\pm 0.12$\\
\hline
Rank-LSTM &$4.17e^{-2}\pm 7.50e^{-3}$ &$0.45\pm0.27$&$ 1.10\pm 0.77$\\
\hline
RSR &$\underline{5.07e^{-2}\pm 7.09e^{-3}}$ &$\underline{0.68\pm0.60}$&$ \mathbf{1.55\pm 0.82}$\\
\hline
STHAN &$3.17e^{-2}\pm 5.09e^{-3}$ &$0.44\pm0.53$&$ 1.42\pm 1.07.$\\
\hline
FinGAT&$4.80e^{-2}\pm 7.85e^{-3}$ &$0.51\pm0.43$&$ 1.25\pm 0.88$\\
\hline
\hline
H-GAT&$\mathbf{7.70e^{-2}\pm 4.33e^{-3}}$ &$\mathbf{0.85\pm 0.59}$&$\underline{1.53\pm 0.75}$\\
\hline
\hline

\multirow{2}*{\ }& \multicolumn{3}{c}{\textbf{NYSE}}  \\
\cline{2-4}
& \textbf{MRR}& \textbf{IRR}& \textbf{SR}\\
\hline
Vanilla LSTM &$2.75e^{-2}\pm 1.09e^{-3}$ &$-0.90\pm0.73$&$ 0.12\pm 0.10$\\
\hline
SFM &$4.82e^{-2}\pm 4.95e^{-3}$ &$0.49\pm0.47$&$ 0.96\pm 0.82$\\
\hline
GCN &$\underline{5.01e^{-2}\pm 5.56e^{-2}}$ &$0.79\pm0.56$&$ 1.71\pm 0.92$\\
\hline
Rank-LSTM&$3.79e^{-2}\pm 8.82e^{-3}$ &$0.56\pm0.68$&$ 0.98\pm 0.85$\\
\hline
RSR &$4.51e^{-2}\pm 2.41e^{-3}$ &$\underline{0.96\pm 0.47}$&$\underline{1.82\pm 1.13}$\\
\hline
STHAN &$2.80e^{-2}\pm 4.31e^{-3}$ &$0.33\pm0.22$&$ 1.12\pm 0.96$\\
\hline
FinGAT &$2.8e^{-2}\pm 7.73e^{-3}$ &$0.84\pm0.90$&$ 1.34\pm 0.98$\\
\hline
\hline
H-GAT&$\mathbf{7.54e^{-2}\pm 5.11e^{-3}}$ &$\mathbf{1.01\pm0.78}$ & $\mathbf{1.97\pm1.12}$\\
\hline
\hline
\end{tabular}
\caption{Performance comparison with baseline models on Nasdaq and NYSE stock markets. Bold \& underlines indicate the first \& second best models, respectively. H-GAT beats baseline models in MRR and IRR evaluation metrics, and only had a slightly lower Sharpe Ratio than RSR on Nasdaq.}
\label{performance}
\end{table*}

\subsection{Baselines}
We compare H-GAT with the following seven state-of-the-art methods:

\begin{itemize}
\item \textbf{Vanilla LSTM}~\cite{bao2017deep}: It  forecasts stock index trends, taking into account short - and long-term effects, and has performed well in six stock markets in different countries.
\item \textbf{State Frequency Memory Recurrent Network (SFM)}~\cite{zhang2017stock}: The SFM decomposes stock prices into different frequencies using Discrete Fourier Transform (DFT).  Low frequency focuses on long-term returns and high frequency focuses on short-term returns. Then feed frequency coefficients to LSTM to learn the sequence characteristics and predict the stock price.
\item \textbf{Graph Convolutional Network (GCN)}~\cite{kipf2017semi}: The GCN is a classical graph-base learning model in spectral domain. The GCN layer calculates symmetric normalized Laplacian matrix from adjacency matrix, then aggregates features of adjacent nodes through graph Fourier transform.
\item \textbf{Rank-LSTM}~\cite{feng2019temporal}: It improves vanilla LSTM loss function by introducing pair-wise loss and excludes stock relations.

\item \textbf{Relational Stock Ranking (RSR)}~\cite{feng2019temporal}: It proposes graph neural network to model the relationship between stocks, and design TGC component to propagate node characteristics in the graph in a temporal-aware manner.
\item \textbf{Spatio-temporal Hypergraph Attention Network (STHAN)}~\cite{sawhney2021stock}: This model contributes to the use of the same relationship to build the hyperedge, and then extend the stock graph from ordinary graph to hypergraph\footnote{The results of STHAN are from the original paper.}.
\item \textbf{Financial Graph Attention Network (FinGAT)}~\cite{ma2022fuzzy}: The FinGAT breaks through the limitation of ordinary graph, constructs the correlation matrix of hypergraph by fuzzy clustering, and reasonably models the potential high-order relationship between stocks.
\end{itemize}

\subsection{Profitability Study (RQ1)}

The Table \ref{performance} shows the performance comparison on NASDAQ and NYSE. We have the following observations:
\begin{itemize}
\item Our H-GAT achieves the best performance over five out of eight settings.
\item From the perspective of investment, our H-GAT has the best IRR in both datasets, which indicates that it achieves the highest profits. Specifically, our H-GAT is significantly increased by 25.00\% and 5.20\% over the second best method RSR on both datasets.
\item For the evaluation of MSE, the performance of our H-GAT is inferior to SFM since the loss of the H-GAT balances between MSE and ranking. By contrast, the SFM only focuses on the MSE loss.
\end{itemize}
We assume that our model H-GAT benefits from the higher-order structures and FA factors from financial statements. In Sec. \ref{sec:HO} and Sec.\ref{sec:FA},  we will perform ablation analysis to further verify their effectiveness.

\subsection{Effectiveness of Individual Higher-order Structure (RQ2)} \label{sec:HO}
This section aims to study how different higher-order structures affect the performance of our H-GAT. Following the below Eq.(\ref{eq:HD}), we first calculate the density of higher-order structure in graphs of NASDAQ and NYSE as follows:

\begin{equation} \label{eq:HD}
    d_{M_{i}}=\frac{|\mathbf{{\boldsymbol\zeta}_{m}}(M_{i})|}{|\mathbf{{\boldsymbol\zeta}_{m}}|},
\end{equation}
where $|\mathbf{{\boldsymbol\zeta}_{m}}(M_{i})|$ indicates $n$-node subgraph sets in which are isomorphic to motif $M_{i}$, $\mathbf{{\boldsymbol\zeta}_{m}}$ indicates all $n$-node subgraphs in graph $G$. In Table \ref{Motif RPD}, we show the relative performance decrease (RPD) and the HoS density. The RPD quantifies the performance reduction if the one type of HoS is not incorporated into our model. We have the following observations:

\begin{itemize}
\item The number of HoS in both datasets are unbalanced. For example, the percentage of the type M4 consisting of three interrelated entities is more than 98\% on both datasets. By contrast, the percentage of the type M3 is only about 1\%. In addition, some HoSs (e.g., M1, M2, M5, M6, M7) are not found in both datasets.
\item The the incorporation of the HoS can improve the performance of H-GAT. For example, the performance of H-GAT is decreased by 36.10\% and 24.47\% on both datasets if it does not incorporate M4. It is noting that the M13 only take 1.51\% over all types of the HoS on NASDAQ, but the performance of our model will be significantly affected 60.27\%. You can have a similar scenario on NYSE. 
\end{itemize}

\begin{table}[t]  
\centering 
\subtable[NASDAQ]{
\begin{tabular}{|m{0.03\textwidth}|m{0.06\textwidth}<{\centering}|m{0.06\textwidth} <{\centering}|m{0.065\textwidth}<{\centering}|m{0.065\textwidth}<{\centering}|m{0.06\textwidth}<{\centering}|}
\hline
\multirow{2}*{HoS}&M3 & M4&M8&M10&M13 \\
\cline{2-6}
\rule{0pt}{25pt} &\adjustbox{valign=c}{\includegraphics[width=0.05\textwidth, height=0.05\textwidth]{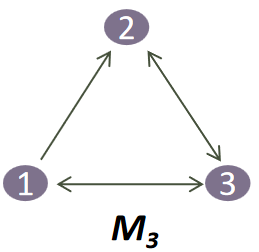}}
& \adjustbox{valign=c}{\includegraphics[width=0.05\textwidth, height=0.05\textwidth]{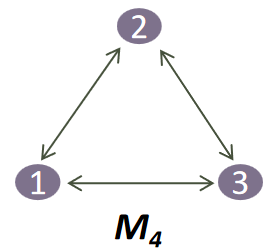}}
&\adjustbox{valign=c}{\includegraphics[width=0.05\textwidth, height=0.05\textwidth]{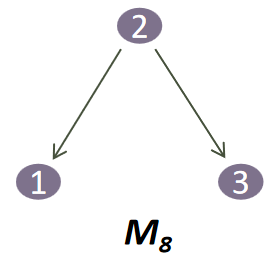}}
&\adjustbox{valign=c}{\includegraphics[width=0.05\textwidth, height=0.05\textwidth]{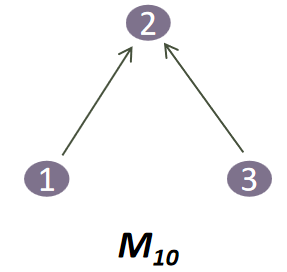}}
&\adjustbox{valign=c}{\includegraphics[width=0.05\textwidth, height=0.05\textwidth]{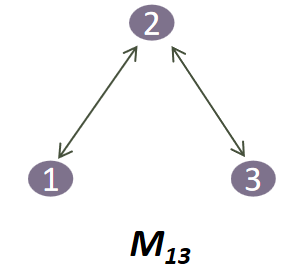}} \\
\hline
$d_{M_{i}}$ (\%)  &0.01 & 98.47&2.86e-05&2.86e-05&1.51\\
\hline
RPD (\%)&14.38&36.10&-1.20&-1.20&60.27\\
\hline
\end{tabular}
}
\subtable[NYSE]{
\begin{tabular}[t]{|m{0.03\textwidth}|m{0.045\textwidth}<{\centering}|m{0.05\textwidth}<{\centering}|m{0.05\textwidth}<{\centering}|m{0.05\textwidth}<{\centering}|m{0.05\textwidth}<{\centering}|m{0.045\textwidth}<{\centering}|}
\hline
\multirow{2}*{HoS}&M3 & M4&M8&M10&M11&M13 \\
\cline{2-7}
\rule{0pt}{25pt} &\adjustbox{valign=c}{\includegraphics[width=0.05\textwidth, height=0.05\textwidth]{m3.png}}
& \adjustbox{valign=c}{\includegraphics[width=0.05\textwidth, height=0.05\textwidth]{m4.png}}
&\adjustbox{valign=c}{\includegraphics[width=0.05\textwidth, height=0.05\textwidth]{m8.png}}
&\adjustbox{valign=c}{\includegraphics[width=0.05\textwidth, height=0.05\textwidth]{m10.png}}
&\adjustbox{valign=c}{\includegraphics[width=0.05\textwidth, height=0.05\textwidth]{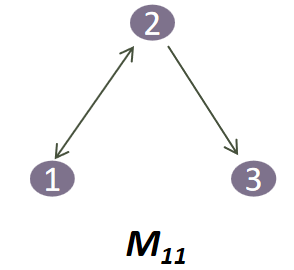}}
&\adjustbox{valign=c}{\includegraphics[width=0.05\textwidth, height=0.05\textwidth]{m13.png}} \\
\hline
$d_{M_{i}}$ (\%)  &0.01 & 98.19&1.73e-05&1.73e-05&1.73e-05&1.79\\
\hline
RPD (\%)& 54.00&24.47&0.77&0.77&0.77&85.14\\
\hline
\end{tabular}
}
\caption{Impacts of removing specific motif regarding IRR's relativly performance decrease (RPD)} 
\label{Motif RPD}
\end{table}

\subsection{Effectiveness of FA Factors and All Higher-order Structures  (RQ3)} \label{sec:FA}

 

\begin{figure}[t]
\centering
\begin{multicols}{1}
\includegraphics[width=85mm]{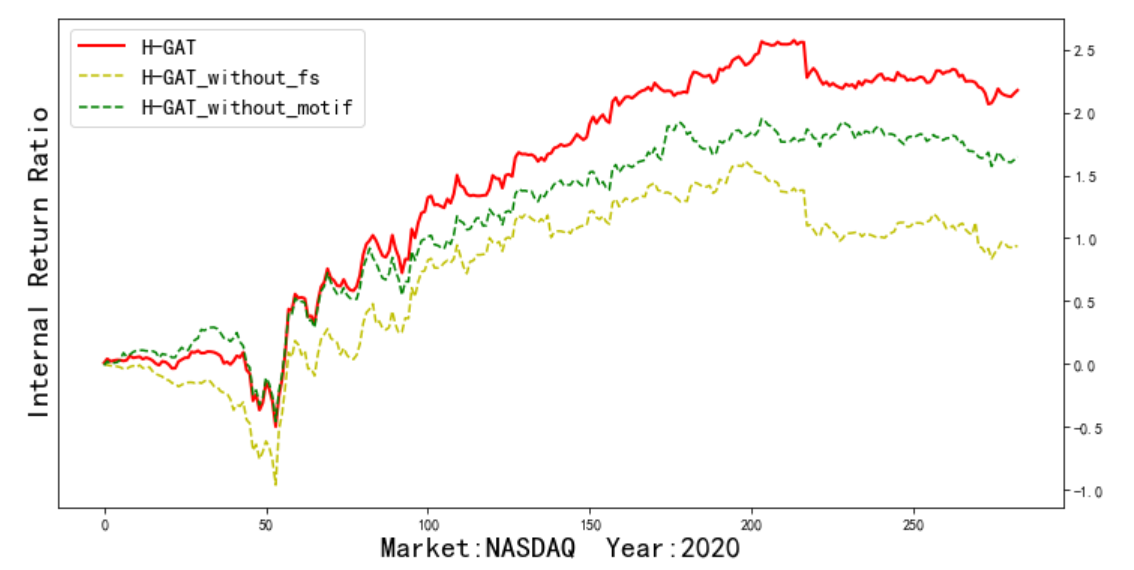}
\end{multicols}
\begin{multicols}{1}
\includegraphics[width=85mm]{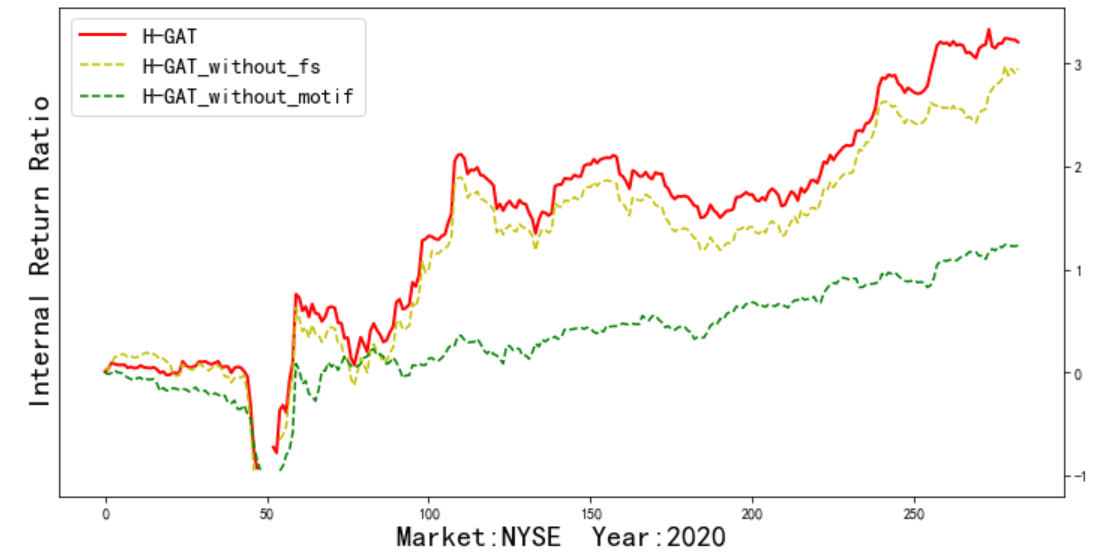}
\end{multicols}
\caption{Ablation test. Both financial statements and network motifs can improve model performance. }
\label{ablation}
\end{figure}

In this section, we study the effectiveness of FA factors shown in Table \ref{financial features}. The Fig.\ref{ablation} shows the performance of the H-GAT that does not incorporate FA and HoS on IRR over the NASDAQ and NYSE in the year of 2020. We have the following observations:

\begin{itemize}
\item The performance of H-GAT is significantly improved if it incorporates both FA factors and all higher-order structures.
\item The HoS is more beneficial to H-GAT on the dataset NYSE than the dataset NASDAQ. The reason would be that the density of graph of NYSE is higher that that of NASDAQ. 
\end{itemize}
 

\subsection{Hyper-parameter Sensitivity Analysis}
\begin{figure}[t]
\begin{multicols}{2}
    \includegraphics[width=\linewidth]{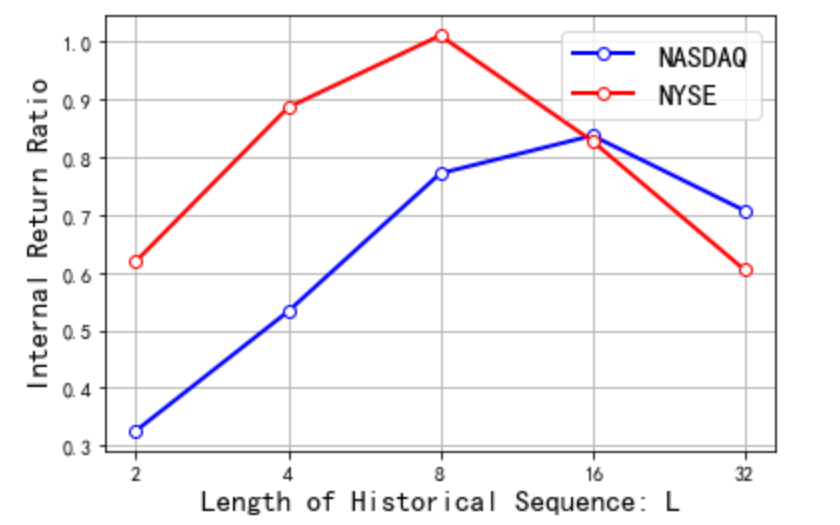}\par 
    \includegraphics[width=\linewidth]{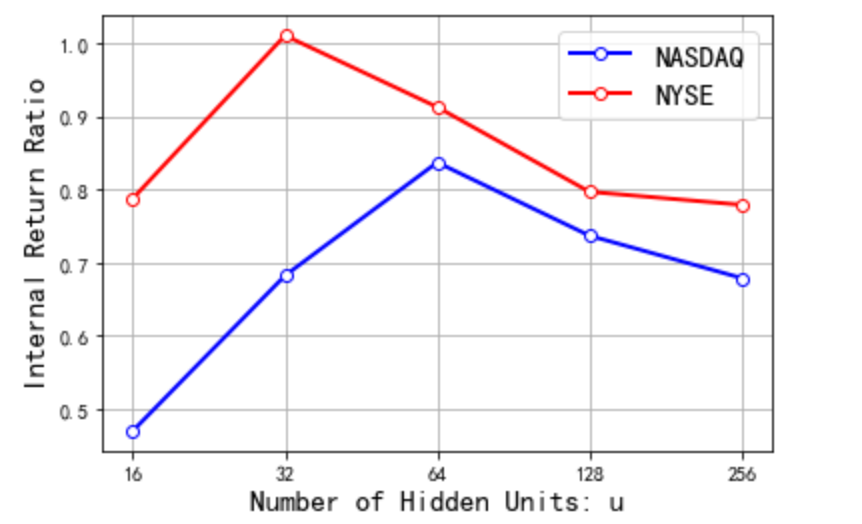}\par 
    \end{multicols}
    \begin{multicols}{2}
    \includegraphics[width=\linewidth]{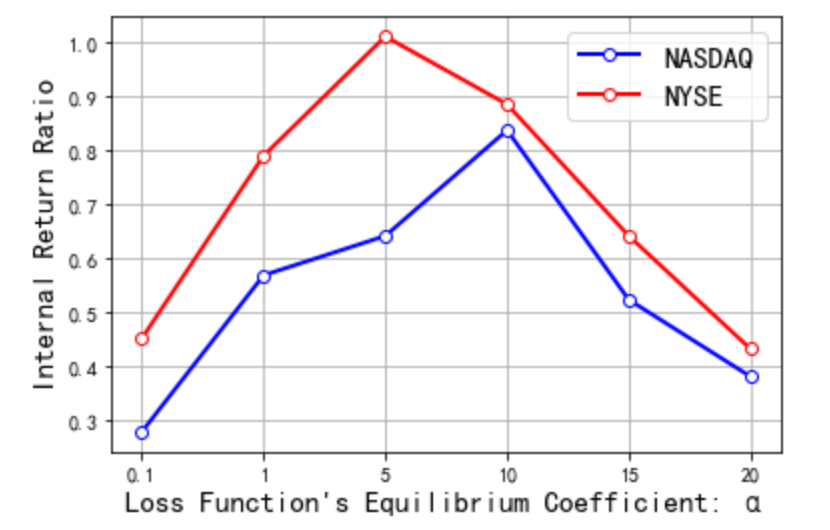}\par 
    \includegraphics[width=\linewidth]{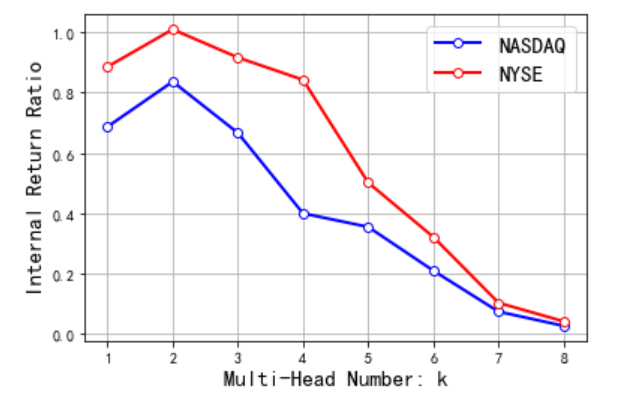}\par
    \end{multicols}
\caption{Sensitivity to hyper-parameters l,u,$\alpha$ and k.}
\label{fig:fa_factors}
\end{figure}


In this section, we conduct sensitivity analysis for four hyper-parameters that are length of historical sequence \textit{l}, the number of LSTM hidden units \textit{u}, the equilibrium coefficient of loss function \textit{$\alpha$} and the number of multi-head \textit{k}. From the Fig.\ref{fig:fa_factors}, we have the following observations:

\begin{itemize}
\item The performance of the model improves gradually with the increases of $l$ because more information is being incorporated into our model. However, for NASDAQ and NYSE, their performance reaches a maximum point at sequence lengths of 16 and 8, respectively, and then decreases slowly.
\item The increase in the number of hidden units ($u$), allows deep learning models to learn more, which is the main reason why model performance initially improves. Nevertheless, larger values of $u$ do not always lead to better performance. When $u$ exceeds 32 or 64, the model tends to overfit and this can cause training time to increase.
\item The impact of the equilibrium coefficient($\alpha$), on the IRR differs significantly depending on the exchange. In the case of NASDAQ, the H-GAT approach achieves the highest performance when the ratio of regression loss to ranking loss is 1:10. On the other hand, NYSE reaches its peak performance when the ratio is 1:5. The reason for this difference is that NASDAQ tends to be more unstable, making it crucial to place greater emphasis on the relative ranking order of the losses.
\item Multi-head attention is designed to attend to different feature spaces, and the selection of the number of attention heads ($k$), is associated with the feature dimension. The experiment demonstrates that both markets achieve the best performance when $k$ equals 2. This could be attributed to the fact that the feature dimension of the sequential features, including stock price and financial statements, is one-to-one.
\end{itemize}

\begin{table*}[t]
\renewcommand{\arraystretch}{1.2}
\centering
\begin{tabular}{|m{0.15\textwidth}<{\centering}|m{0.1\textwidth}||m{0.07\textwidth}<{\centering}|m{0.07\textwidth}<{\centering}|m{0.07\textwidth}<{\centering}|m{0.07\textwidth}<{\centering}|}
\hline
\multicolumn{2}{|c||}{\multirow{2}{*}{\textbf{Portfolios}}} &\multicolumn{2}{c|}{\textbf{IRR}}&\multicolumn{2}{|c|}{\textbf{SR}}\\
\cline{3-6}
\cline{3-6}
\multicolumn{2}{|c||}{}&NASDAQ &NYSE &NASDAQ &NYSE\\
\hline
\hline
\multirow{3}{*}{H-GAT}&Top 1& \textbf{1.10}& \textbf{1.02} & \textbf{1.70} & \textbf{1.59}\\
\cline{2-6}
&Top 5& 0.49 &0.47 &1.24 &1.03\\
\cline{2-6}
&Top 10& 0.33& 0.35 & 0.96 &0.87\\
\hline
\multicolumn{2}{|c||}{Standard and Poor's 500} & \multicolumn{2}{c|}{0.12}&\multicolumn{2}{|c|}{0.98} \\
\hline
\multicolumn{2}{|c||}{Dow Jones Industrial Average} & \multicolumn{2}{c|}{0.11}&\multicolumn{2}{|c|}{0.88}\\
\hline
\end{tabular}
\caption{Portfolio performance on test days: 01/03/2017 - 07/05/2022}
\label{tab}
\end{table*}

\begin{figure*}[t] 
    \center{\includegraphics[width=0.7\linewidth]{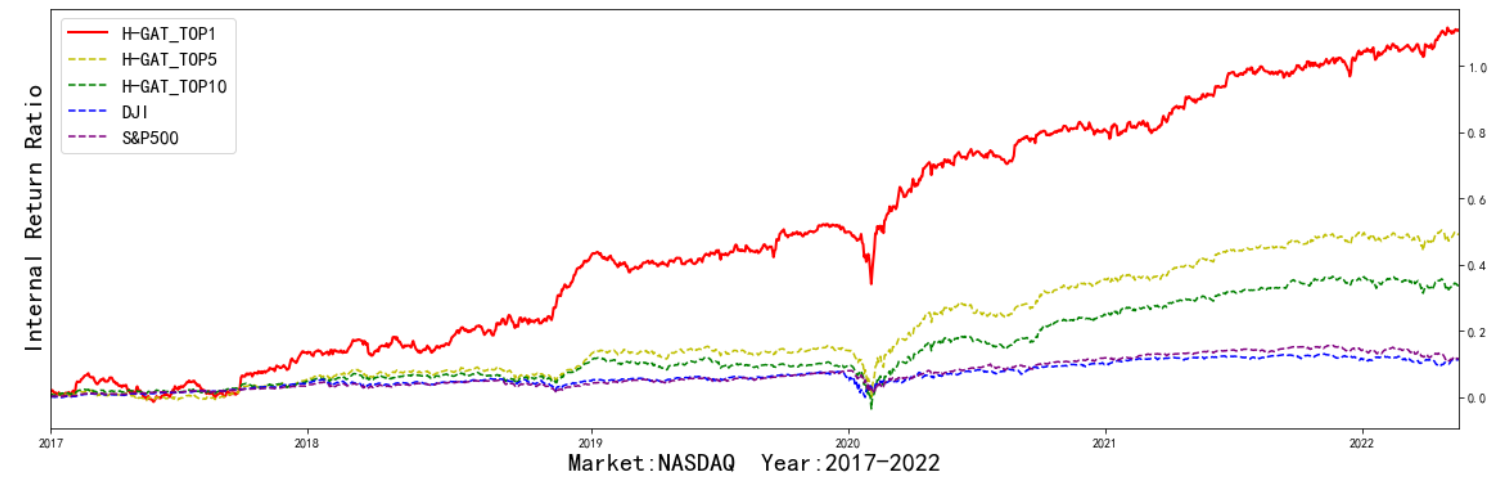}}
\caption{Portfolios' Internal Return Ratio of H-GAT over a long-term span on NASDAQ for a profitability analysis. The experimental results on NYSE are very similar with it. } \label{tab:port}
\end{figure*}

\subsection{Long-term Profitability Analysis (RQ4)}
   Most papers only show a very short time-span to test the profitability, which leads to consider either bull or bear market, not both. The paper~\cite{gonzalez2005two} indicates that the average bull market in U.S. is nearly 21 months long, while the average bear market is 15 months long. We extend time span and choose Top N (i.e., 1, 5, 10) stocks to examine the performance of H-GAT across the bull - bear cycle. Over a backtest period of up to 5 years, we have following observations:
\begin{itemize}
    \item Our model outperforms the Dow Jones Industrial Average (DJI) and the Standard and Poor's 500 (S\&P 500). It indicates that H-GAT is worthy for long-term investment, with an annualized IRR 9-10 times that of the stock index.
    \item When the trading ratio of Top N stocks is equal, the ranking of profitability is Top 1 $>$ Top 5 $>$ Top 10. This suggests that taking buy-hold-sell multiple stocks in the top order cannot effectively share risks, but will reduce the returns of the model. In the future, we can consider combining short selling mechanism -- 'sell, hold and buy' to offset risks.
    \item Nonetheless, when comes to Sharpe Ratio, the advantage of the H-GAT model is less evident compared with IRR. The Sharpe ratio of the Top 1 on NSADAQ and NYSE is 1.70 and 1.59. The sharp ratio of two stock indexes was 0.98 (S\&P 500) and 0.88 (DJI). This indicates that this approach generates large returns but also carries significant risks. 
\end{itemize}

\section{Conclusion}
In this paper, we construct a new stock dataset and propose H-GAT to optimally select stocks. We use higher-order structures to represent the relation among more than two stocks. The H-GAT model takes into account both temporal attention and functional attention to capture the sequential dependencies within stock graphs. Simultaneously, it extracts higher-order structures from the graphs to learn stock embeddings.
In the future, we will import more risk control mechanisms. For example, industry risk can be reduced by diversification, such as ranking and trading the Top N stocks in each sector separately. 

\bibliography{bibliography.bib}

\end{document}